\documentclass[twocolumn,aps,prl,nofootinbib,nobibnotes,superscriptaddress,tightenlines]{revtex4-1}

\usepackage{amsmath, amsfonts, amsthm, amssymb, graphicx, color, hyperref}
\usepackage{subfigure}

\usepackage{slashed}
\usepackage{physics}
\usepackage{epstopdf}

\allowdisplaybreaks[3]

\begin{document}

\hfill{\footnotesize USTC-ICTS/PCFT-20-34}

\title{Gravitational collapse with quantum fields}

\author{Benjamin Berczi}
    \email[]{benjamin.berczi@nottingham.ac.uk}
    \affiliation{School of Physics and Astronomy, University of Nottingham, University Park, Nottingham NG7 2RD, United Kingdom}
\author{Paul M. Saffin}
    \email[]{paul.saffin@nottingham.ac.uk}
    \affiliation{School of Physics and Astronomy, University of Nottingham, University Park, Nottingham NG7 2RD, United Kingdom}
\author{Shuang-Yong Zhou}
    \email[]{zhoushy@ustc.edu.cn}
    \affiliation{Interdisciplinary Center for Theoretical Study, University of Science and Technology of China, Hefei, Anhui 230026, China}
    \affiliation{Peng Huanwu Center for Fundamental Theory, Hefei, Anhui 230026, China}

\date{\today}

\begin{abstract}

Gravitational collapse into a black hole has been extensively studied with classical sources. We develop a new formalism to simulate quantum fields forming a black hole. This formalism utilizes well-established techniques used for classical collapse by choosing a convenient coherent state, and simulates the matter fields quantum mechanically. Divergences are regularized with the cosmological constant and Pauli-Villars fields. Using a massless spherically symmetric scalar field as an example, we demonstrate the effectiveness of the formalism by reproducing some classical results in gravitational collapse, and identifying the difference due to quantum effects. 

\end{abstract}

\maketitle

{\bf Introduction~}
The simplest model of a black hole forming in 4D is the collapse of a 
massless scalar field. The classical aspects of scalar collapse have been extensively studied in spherical symmetry analytically by Christodoulou \cite{christodoulou1, christodoulou2, christodoulou3, christodoulou4} and  numerically in \cite{goldpiran} and also \cite{Choptuik} where Choptuik famously uncovered the phenomenon of critical collapse that exhibits a range of fascinating properties \cite{roberts1989scalar, Brady_1994, Oshiro_1994, frolov1999self}.

The semi-classical analysis by Hawking \cite{hawkingradiation} shed new light on the quantum behavior of black holes, discovering the celebrated Hawking radiation. However, non-linearly solving scalar collapse semi-classically is a tremendous task, as the quantum evolution of matter fields with full gravitational effects is difficult to solve even in spherical symmetry. Early attempts towards investigating quantum scalar collapse were made in \cite{tomimatsu, Bak:1999wb, Bak:2000kg}, as well as studying the quantum behaviour of a collapsing spherical shell \cite{Vachaspati:2007hr, Greenwood:2008ht}, and the modelling of the quantum radiation with classical fields \cite{Vachaspati:2018pps}. Numerical evidence for Hawking radiation has been found in analogue systems \cite{,Balbinot:2007de, Carusotto:2008ep}. Simpler models such as the collapse of a dilaton in 2D gravity have been studied \cite{Russo:1992ax, Strominger:1993tt, Piran:1993tq, Lowe:1992ed}, but these do not contain the spherical harmonics that originate in the 4D system.

In this letter, we establish a new method to numerically study the collapse of a fully quantum source of matter into a black hole. This is done by coupling a quantum scalar field to classical gravity and numerically simulating its evolution towards gravitational collapse. By choosing a convenient coherent state, the gravitational source is naturally split into the classical contribution plus the quantum fluctuations, which allows us to modify the well-studied classical collapse to achieve a quantum simulation. This method has many applications. For example, it can be used to investigate how quantum effects change the universality and scaling of classical critical collapse, and also to simulate Hawking radiation in real time. 

We have focused on the simplest model of a massless scalar field, but this formalism is applicable to generic free quantum fields, including fermionic fields, which are inherently quantum in nature. Some initial results will be presented in this letter, which are comparable to those of a classical collapse found in, for example, \cite{alcubierre} and from which we can already identity quantum corrections in gravitational collapse. More details of the method, along with its applications, will be presented elsewhere \cite{futurepaper}. 

{\bf Classical collapse~}
The main idea is to promote the classical scalar involved in gravitational collapse to a quantum field operator. Before that, let us briefly review the classical collapse. Our notation and method follow that of \cite{alcubierre}. The metric in spherical symmetry is chosen to be
\begin{equation} \label{linelement}
    ds^2=-\alpha^2(t,r)dt^2+A(t,r)dr^2+r^2B(t,r)d\Omega^2,
\end{equation}
where $d\Omega^2=d\theta^2+\sin^2\theta d\phi^2$. The stress-energy tensor of the classical massless scalar field $\Phi(t,r)$ is
\begin{equation}
    T_{\mu \nu} = \partial_{\mu}\Phi \partial_{\nu}\Phi 
                    -\frac{1}{2} g_{\mu \nu}\Big[g^{\rho\sigma} \partial_{\rho}\Phi \partial_{\sigma}\Phi 
                     \Big].
\end{equation}
The Einstein equations and the scalar field equation, respectively, are
\begin{align} 
\label{fieldeqs}
    M_P^2 G_{\mu \nu} &=  \big(T_{\mu \nu}-\Lambda g_{\mu \nu}\big), \\
\label{fieldeqs1}
    \Box \Phi &= 0,
\end{align}
where $\Lambda$ is the cosmological constant, taken to be zero in the classical case but needed later to regularize divergences in the quantum case \footnote{When performing the simulations with a quantum source, we replace $T_{\mu\nu}$ with $\langle\hat T_{\mu\nu}\rangle$.}. The Einstein equations can be used to find a set of well-behaved first order evolution equations for the metric fields $\big(\alpha(t,r), \: A(t,r), \: B(t,r)\big)$, or rather re-formulated variables with these fields, as done in \cite{alcubierre}.

An important ingredient in the classical collapse is the gauge choice of the lapse function $\alpha(t,r)$.
In \cite{alcubierre}, maximal slicing is used, but this requires integrating spatially an ODE at each time step, which is numerically expensive. We instead use the ``1+log'' gauge \cite{BonaMasso}, which results in $\alpha(t,r)$ becoming a dynamical variable and is nowadays more commonly used due to its simplicity and robustness.

{\bf Quantization~}
In the quantum simulations,  the scalar field is promoted to be an operator and can be expanded in terms of spherical mode functions as
\begin{equation} \label{mode_expansion}
    \hat{\Phi}=\sum_{l,m} \int dk [\hat{a}_{k,l,m}\Tilde{u}_{k,l}(t,r)Y^m_l(\theta, \varphi) + h.c.],
\end{equation}
where $Y^m_l(\theta, \varphi)$ are the spherical harmonics and $\Tilde{u}_{k,l}(t,r)=r^l u_{k,l}(t,r)$ are the quantum mode functions, which in Minkowski spacetime are just the spherical Bessel functions $\Tilde{u}_{k,l}(t,r)=\frac{k}{\sqrt{\pi \omega}} e^{-i\omega t} j_l(kr)$. The operator $\hat{a}_{k,l,m}$,  is a ladder operator that satisfies $\left[\hat{a}_{k,l,m},\hat{a}_{k,l,m}^\dagger \right] = \delta_{ll'}\delta_{mm'}\hbar c^2\delta(k-k')$.

Also, the quantum state is chosen to be a particular coherent state
\begin{align}
    \ket{\chi}&=\exp\left[-\frac{1}{2}\frac{1}{\hbar c^2}\int dk \abs{z(k)}^2\right] \nonumber\\ 
    &\quad \quad~~~~ \cdot \exp\left[\frac{1}{\hbar c^2}\int  dk z(k) \hat{a}^{\dag}_{k,0,0}\right] \ket{0}, 
\end{align}
where the first exponential is to set $\langle\chi|\chi\rangle=1$. Coherent states are ``minimal uncertainty states'' (see {\it e.g.}~\cite{Sanders_2012}), which are useful to relate quantum systems to classical systems, and this will allow us to make a direct comparison with classical simulations. For this coherent state we have the standard property that it is an eigenstate of the lowering operator,
\begin{equation}
    \hat{a}_{k,0,0} \ket{\chi} = z(k) \ket{\chi}.
\end{equation}
In addition, we define the expectation value of the field operator $\hat{\Phi}$ in the coherent state $\ket{\chi}$
\begin{equation}
    \phi (t,r) = \bra{\chi} \hat{\Phi} \ket{\chi},
\end{equation}
which we note is spherically symmetric and may be written as
\begin{equation}
    \phi (t,r) = \frac{1}{2 \sqrt{\pi}} \int dk [z(k) \Tilde{u}_{k,0} (t,r) + h.c.].
\end{equation}
Note that $\phi(t,r)$ and the classical field $\Phi(t,r)$ satisfy the same equation of motion. This crucial fact allows us to tap into the established method for classical scalar collapse, and simply add the quantum modes on top of the classical fields. Specifically, the evolution equations of the quantum mode functions can be found using the operator equation $\Box \hat \Phi = 0$ and the operator expansion \eqref{mode_expansion} so that:
\begin{align} \label{pi_mode}
    \partial_t{\pi}_{k,l} &= \partial_r\Bigg( \frac{\alpha B}{A^{\frac{1}{2}}} \Bigg) \Bigg(\frac{l}{r}u_{k,l}+\psi_{k,l} \Bigg)  \nonumber \\
    &+ \frac{\alpha B}{A^{\frac{1}{2}}}
    \Bigg( \frac{2l+2}{r}\psi_{k,l} +\partial_r \psi_{k,l} \Bigg) \nonumber\\
    &- \alpha A^{\frac{1}{2}}\Big( \frac{B}{A}-1 \Big)\frac{l(l+1)}{r^2} u_{k,l},
\end{align}
where $\pi_{k,l} = (A^{\frac{1}{2}}B/ \alpha) \partial_t{u}_{k,l}$ and $\psi_{k,l}= \partial_{r} u_{k,l}$. 
Then the expectation values of the quantum stress-energy tensor $\hat T_{\mu\nu}$ are nicely split into the ``classical'' coherent part plus the terms due to quantum fluctuations. For example, the two time derivative bilinear term appearing in the expectation value of the stress-energy tensor can be written as 
\begin{equation} \label{stresscomps}
      \bra{\chi} \partial_t \hat{\Phi} \partial_t \hat{\Phi} \ket{\chi} = \partial_t \phi \partial_t \phi + \frac{\hbar c^2}{4 \pi}\int dk \sum_l (2l+1) \abs{\partial_t \Tilde{u}_{k,l}}^2.
\end{equation}
 We emphasize that this convenient separation is due to our choice of the coherent state, and the scalar field is treated fully quantum mechanically. 

Note that the sums appearing in the stress-energy tensor such as in Eq.~\eqref{stresscomps} are divergent. These need to be regularized, which can be done by introducing a non-zero cosmological constant and five Pauli-Villars fields \cite{paulivillars, Visser:2016mtr, Kamenshchik:2018ttr} whose masses satisfy the relations
\begin{align}
m_{2}^{2}+m_{4}^{2}=m_{1}^{2}+m_{3}^{2}+m_{5}^{2} \\
m_{2}^{4}+m_{4}^{4}=m_{1}^{4}+m_{3}^{4}+m_{5}^{4}
\end{align}
Note that in our method these Pauli-Villars fields are also simulated quantum mechanically with the full machinery of mode function evolutions. The technical details of these will be explained in a future paper \cite{futurepaper}.

\begin{figure}
\centering
   \includegraphics[width=0.8\linewidth]{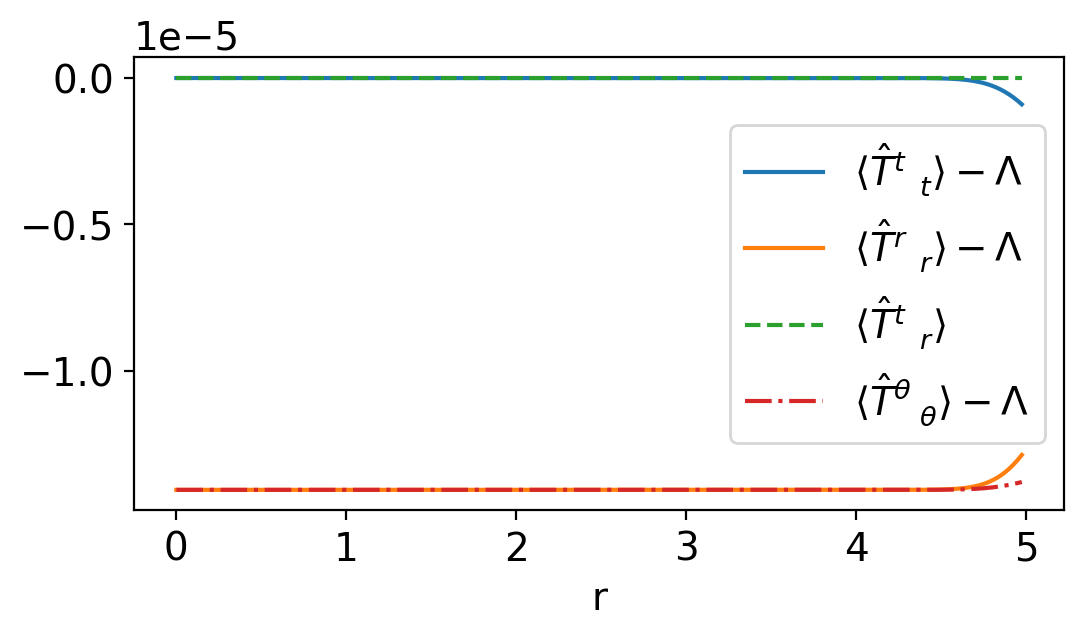}  
   \includegraphics[width=0.8\linewidth]{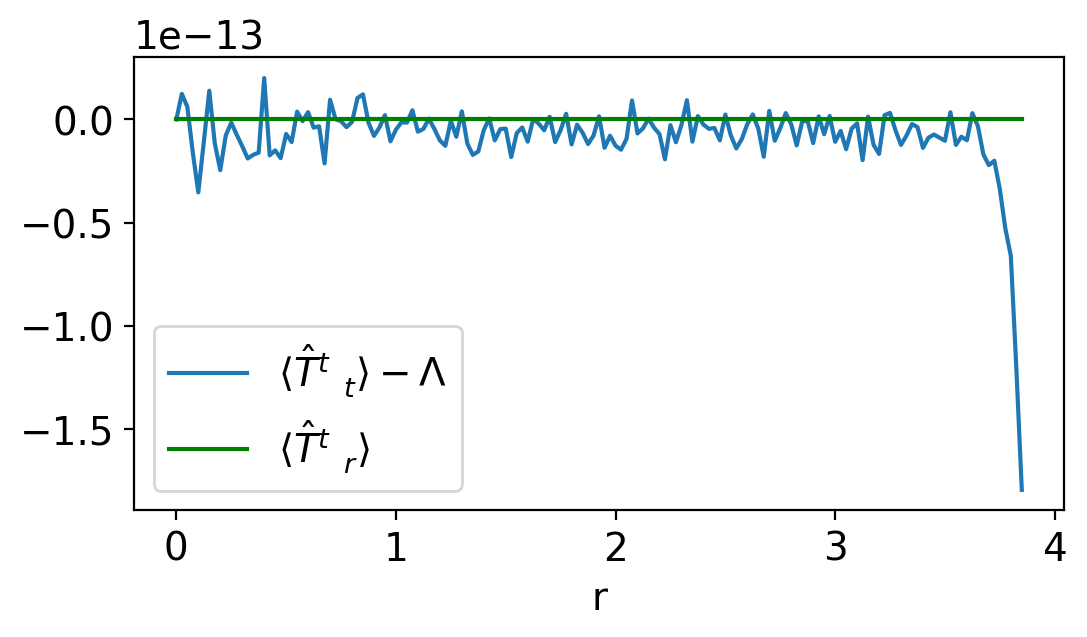}  
\caption{Expectation value of the stress-energy tensor $\langle \hat T^\mu{}_{\nu}\rangle$ in the quantum Minkowski vacuum ($\phi(t,r)=0$) with $N_l=N_k=70$, regularized by five Pauli-Villars ghosts and a cosmological constant. Note that initially $\langle\hat T^t{}_r\rangle$ is exactly zero everywhere since the initial momentum is zero. Our units are such that $M_P=1$.}\label{fig:penrosediag}
\end{figure}

{\bf Numerical setup~}
In the simulation we use a uniform spatial grid consisting of 200 points with $dr=0.025$ and $dt={dr}/{4}$. For spatial derivatives, tenth order finite difference methods are used, and a tenth order implicit Runge-Kutta method \cite{impRK1} is used for time integration. Such high order numerical methods are needed because the evolution equations such as Eq.~\eqref{pi_mode} contain terms involving $1/r, 1/r^2$ and other similar terms, which are numerically unstable if not regulated properly. An alternative route to increase the accuracy would be to make use of the spectral method \cite{futurepaper}. One also needs to use some artificial dissipation, which acts as damping for the numerical errors. In our code, this is done by adding the Kreiss-Oliger terms \cite{dissipation} to each field's evolution equation. Following Alcubierre's notation \cite{alcubierre}, the Kreiss-Oliger term is fourth order with $\epsilon=0.5$ for the quantum mode functions and $\epsilon=0.1$ for the metric fields and $\phi(t,r)$.

\begin{figure}
\centering
   \includegraphics[width=0.8\linewidth]{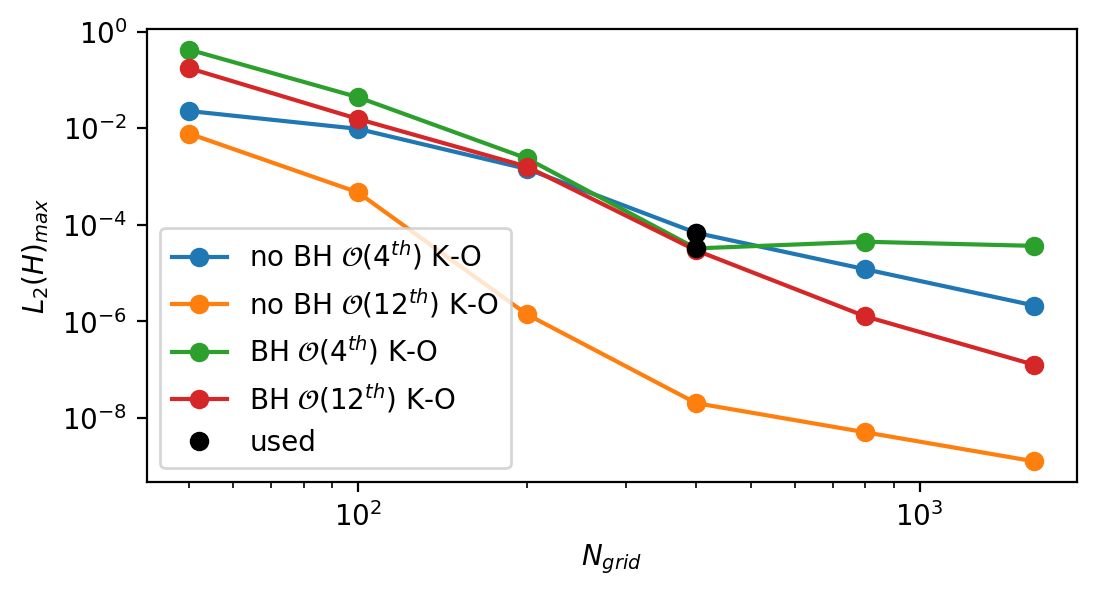}  
   \includegraphics[width=0.8\linewidth]{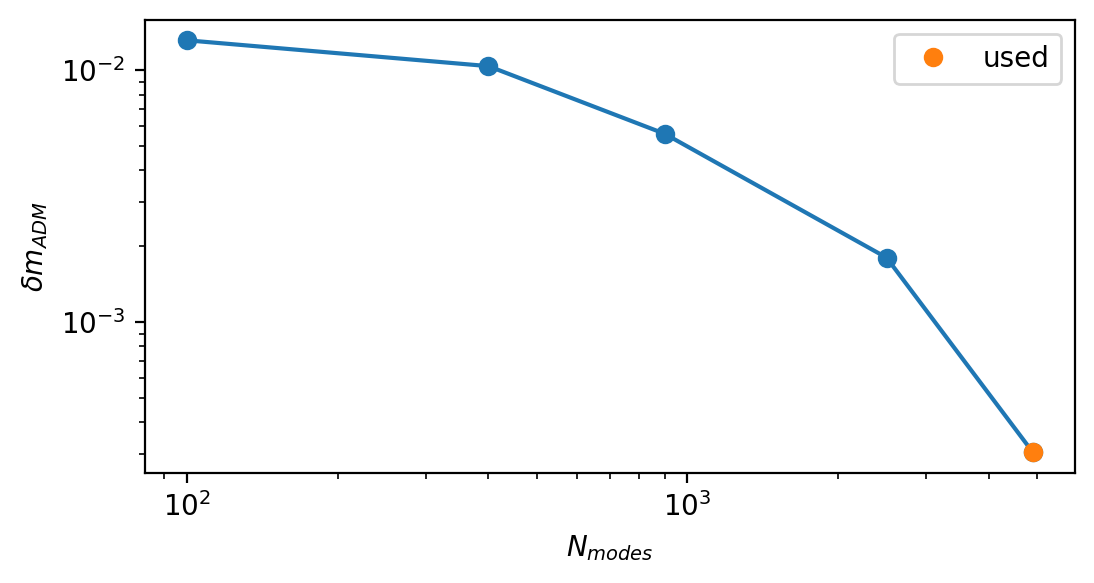}  
\caption{Convergence of the $L_2$ norm of the Hamiltonian constraint with respect to the number of grid points (top) and convergence of the ADM mass with respect to the number of quantum mode functions used (bottom). The dots labeled with ``used'' are the ones that we use in Fig.~\ref{fig:fig} and Fig.~\ref{fig:difference}.}\label{fig:convergence}
\end{figure}

A crucial choice in the simulation is the number of quantum modes $\Tilde{u}_{k,l}$ used, which essentially sets up the quantum field operator. In practice, we can only use a finite number of quantum modes and generally the more quantum modes one uses, the better the simulated quantum operator is. We find that the maximum number of quantum modes one can use is $N_l=N_k=150$, where $N_l$ is the number of $l$-modes and $N_k$ is the number of $k$-modes. This limit arises from the amplitude of the modes becoming  too small, saturating the double precision limit of the simulations. 
In addition, the evolution equations of the quantum modes are increasingly unstable as $N_l$ and $N_k$ increase. A good consistency check is that when the coherent state is the vacuum, $i.e.$ $z(k)=0$, all stress-energy tensor components need to vanish. Using our methods, we have managed to simulate a system with $N_l=N_k=70$, {\it i.e.}, almost five thousand modes, for a sufficiently long time in a sufficiently large spatial region to capture the collapse to a black hole.

Throughout our simulations $M_P$ is set to unity, that is, all the dimensionful quantities are expressed in Planck units. 

\begin{figure*}
\centering
\subfigure[Evolution of $\langle\hat\Phi\rangle$ where a black hole does not form]{
  \includegraphics[width=0.38\linewidth]{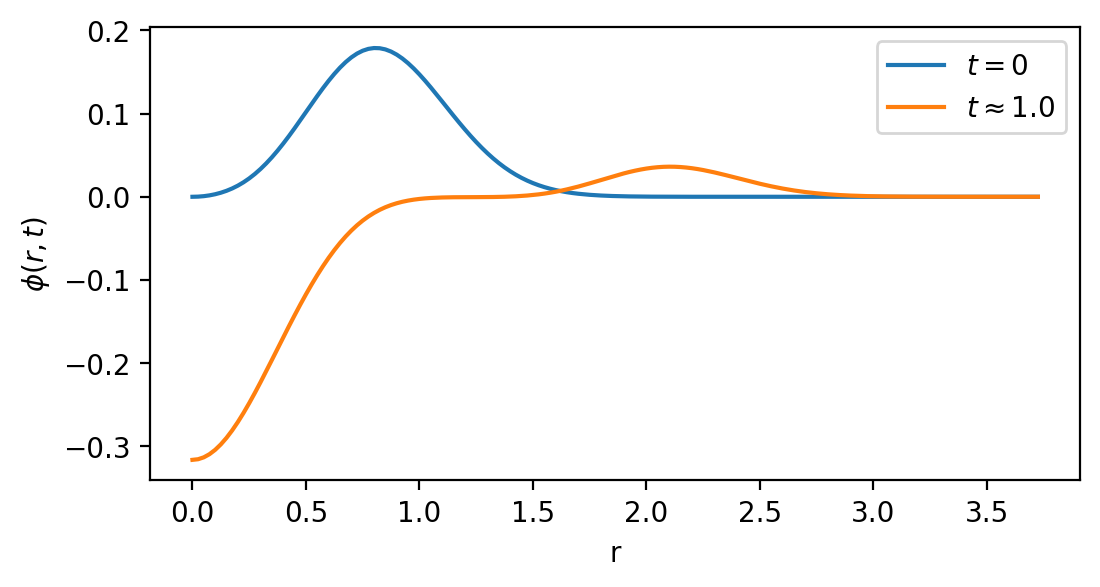} 
}
\subfigure[Evolution of $\langle\hat\Phi\rangle$ where a black hole does form]{
  \includegraphics[width=0.36\linewidth]{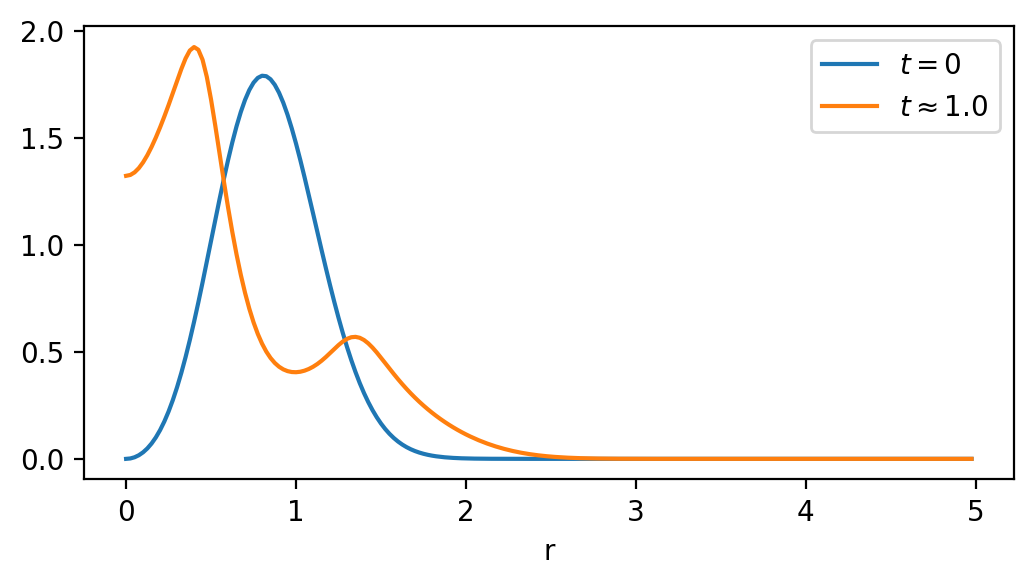}  
} \label{fig:phi2}
\subfigure[Evolution of $\alpha(r=0)$ where a black hole does not form]{
  \includegraphics[width=0.38\linewidth]{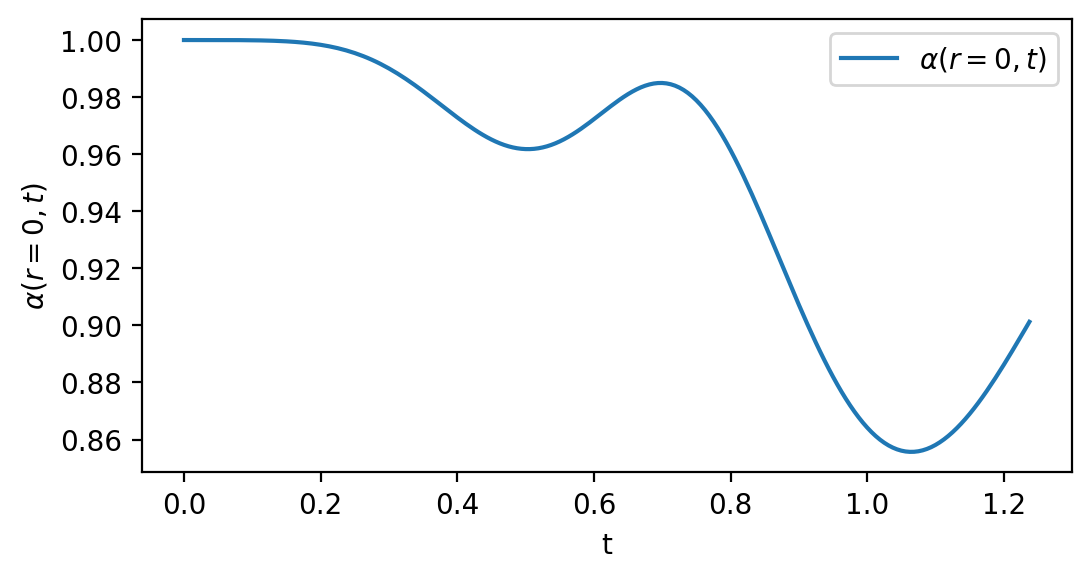}
} \label{fig:alpha1}
\subfigure[Evolution of $\alpha(r=0)$ where a black hole does form, along with the curve showing the area of the apparent horizon.]{
  \includegraphics[width=0.36\linewidth]{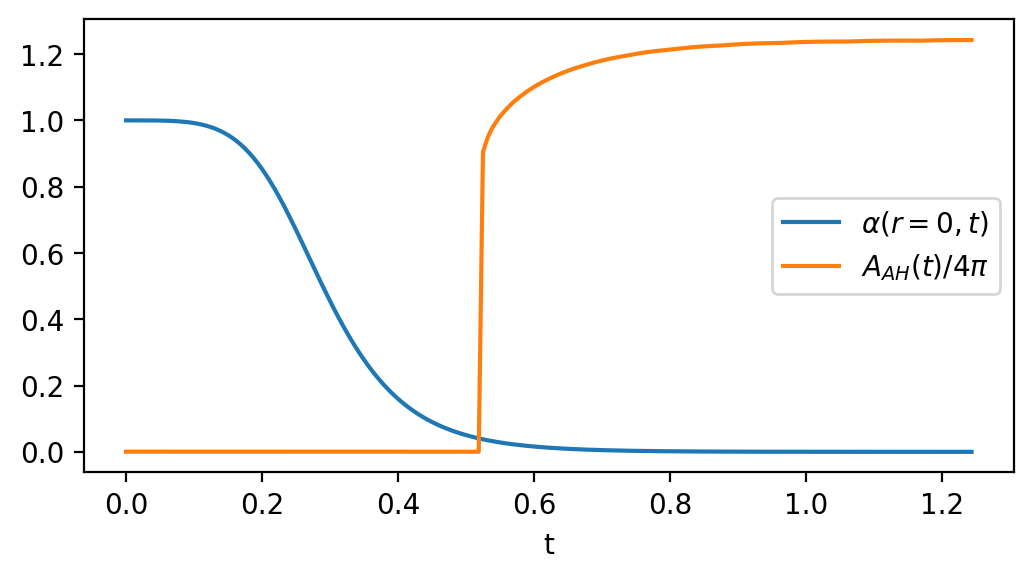}  
}  \label{fig:alpha2}
\subfigure[$\langle \hat{T}^{\mu}_{\: \: \nu}\rangle$ at $t\approx 0.94$ when a black hole does not form]{
  \includegraphics[width=0.38\linewidth]{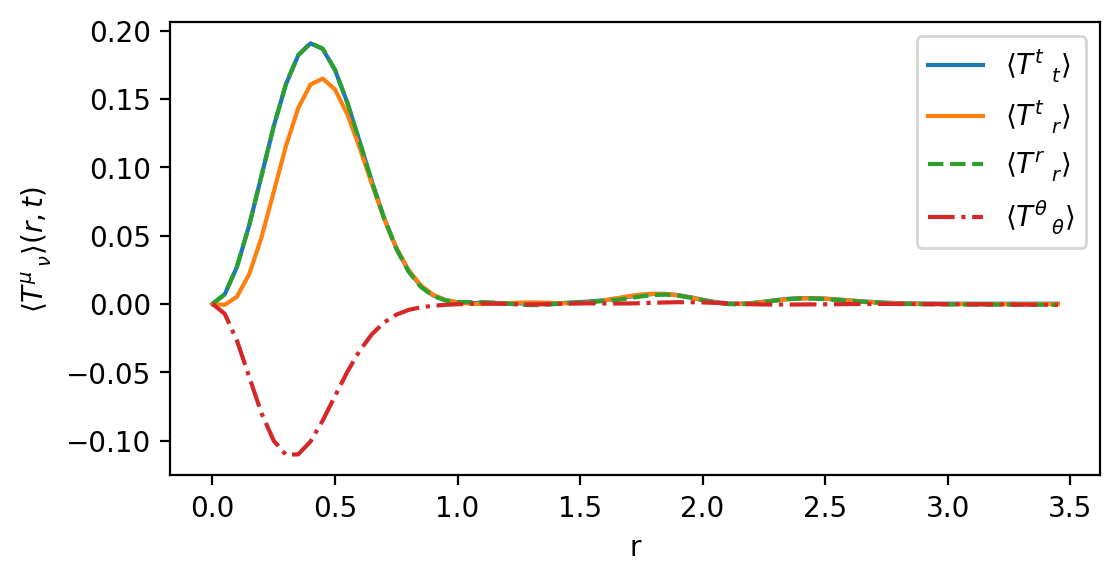}  
}\label{fig:stress1}
\subfigure[$\langle \hat{T}^{\mu}_{\: \: \nu}\rangle$ at $t\approx 0.94$ when a black hole forms]{
  \includegraphics[width=0.36\linewidth]{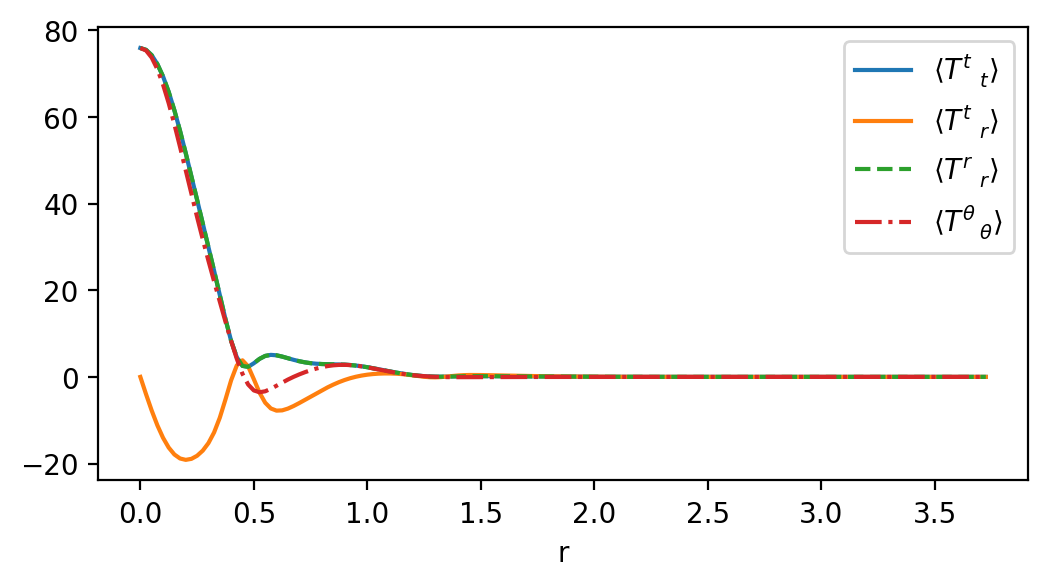}  
}\label{fig:stress2}
\caption{Evolutions of wave-packets with initial amplitude $a=0.1$ (left column) and $a=1.0$ (right column). The latter evolution results in a black hole formation, which can be seen by the collapse of the lapse $\alpha(t,r=0)$ and the presence of an apparent horizon $r_{AH}$.}\label{fig:fig}
\end{figure*}

{\bf Results~} 
\label{sec:conclusions}
In Fig.~\ref{fig:penrosediag}, the initial stress-energy tensor components are illustrated when the coherent state is the vacuum, {\it i.e.}, $\phi(t,r)=0$. This is done with $N_l=N_k=70$ as mentioned above. The cosmological constant is set so that $\langle \hat{T}^t_{\: \: t}\rangle-\Lambda$ equals to zero. However, the stress-energy tensor expectation value is not precisely proportional to the metric, as would be expected in the quantum vacuum, due to the truncation to a finite number of quantum modes, and hence $\langle \hat{T}^r_{\: \: r}\rangle-\Lambda$ and $\langle \hat{T}^{\theta}_{\: \: \theta}\rangle-\Lambda$ are not exactly zero, but under control. Also, the flatness of the stress-energy tensor components breaks down at around $r=5$ before the end of the spatial grid, meaning that the evolution of the fields can only be trusted inside that region.

Initial conditions for the metric fields are derived from the Hamiltonian and momentum constraints, as in \cite{alcubierre}. 
The ``classical'' scalar configuration is initially taken to be $\phi(t=0,r)=f(r)+f(-r)$ and  $\dot\phi(t=0,r)=0$, where $f(r)=a(r/D)^2\exp\left[-((r-R)/D)^2\right]$, with $R=0.5$ and $D=0.5$ and various amplitudes $a$. The quantum modes initially are such that they satisfy Eq.~\eqref{pi_mode} in Minkowski spacetime. For time evolution, we find that the wave-packet splits into two smaller wave-packets travelling in opposite directions, one towards the centre at $r=0$ and one towards the outer boundary. With a sufficiently large initial amplitude the former packet forms a black hole in the centre.

The presence of the black hole is seen in two ways, the first being the collapse of the lapse function $\alpha(t,r)$, and the second being the formation of an apparent horizon. The latter is calculated using the expansion of the outgoing null geodesics $H$, given by
\begin{equation}
    H=\frac{1}{\sqrt{A}}\Bigg( \frac{2}{r} + \frac{\partial_rB}{B} \Bigg) - 2K^{\theta}_{\theta},
\end{equation}
where $K^{\theta}_{\theta}$ is an extrinsic curvature component. If $H$ vanishes for some $r$, then there exists an apparent horizon at that location $r=r_{AH}$, with an area given by $A_{AH}=4\pi r^2_{AH}\left.B\right|_{r_{AH}}$. 

To validate the code, we have performed convergence studies for both the number of points on the grid and the number of quantum modes used, by computing the $L_2$-norm of the Hamiltonian constraint (see e.g. \cite{clough2017scalar}) and the error in the ADM mass (defined by $\Delta m_{ADM}(t)=(m_{ADM}(t=0)-m_{ADM}(t))/m_{ADM}(t=0)$), respectively; see Fig.~\ref{fig:convergence}. The convergence with respect to the number of grid points is shown for 4 different scenarios: evolutions involving black hole/no black hole formation and $4^{th}$ and $12^{th}$ order derivative Kreiss-Oliger terms. 
All the simulations converge, and as expected better convergence can be achieved with $12^{th}$ order Kreiss-Oliger terms for cases without black hole formations. 
%Note that, nevertheless, we use $4^{th}$ order Kreiss-Oliger terms which are necessary due to the unstable numerical nature of the quantum mode functions.
%The latter is included in order to illustrate that the code itself is convergent, even though the black hole case with $4^{th}$ order Kreiss-Oliger term seems to stop converging. This is merely the remedy of the amount of damping used, which is necessary due to the unstable nature of the quantum mode functions.

We present our results in Fig.~\ref{fig:fig} for both no black hole formation (left column) and a black hole formation (right column). These are achieved by using the same parameters (mentioned above) but with different initial amplitudes of $\phi$: $a=0.1$ and $a=1.0$ respectively. The first two rows show the evolution of  $\phi(t,r)=\langle\hat\Phi\rangle$ and $\alpha(t,r=0)$ respectively. In the top left of Fig.~\ref{fig:fig}, one can see that when the amplitude is small, the in-going wave-packet simply gets reflected at the origin. When the amplitude is sufficiently large, a black hole forms. In the top right of Fig.~\ref{fig:fig}, we see that soon after separating both the in-going and out-going wave-packet freeze, indicating a black hole forming, which contains all of the in-going bigger wave-packet and most of the out-going one. In the second row, the collapse of the lapse function $\alpha(t,r)$ can be observed on the right, where the area of the apparent horizon, $A_{AH}$, is also illustrated. The $\langle \hat{T}^{\mu}{}_{\nu}\rangle$ components at $t \approx 0.94$ can be seen in the last row, and we see that the stress-energy tensor is peaked inside the apparent horizon.

\begin{figure}
        \includegraphics[width=0.45\textwidth]{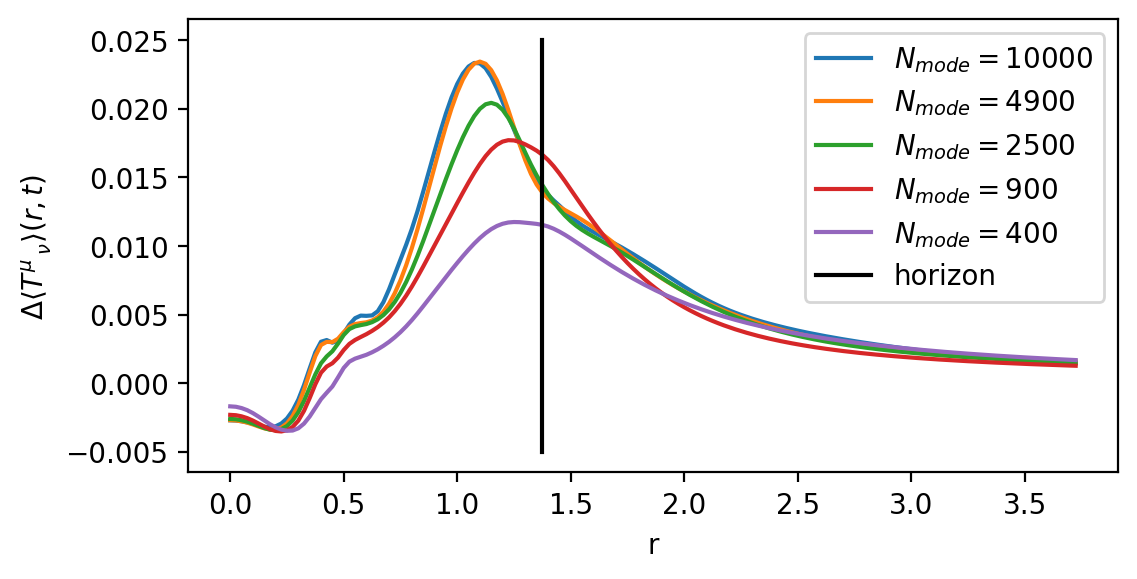}
    \caption{Convergence study of quantum corrections for the stress-energy tensor component $\Delta T^{t}{}_{t} = \bra{\chi}\hat{T}^{t}{}_{t}\ket{\chi} - T^{t}_c{}_{t}$ (with $T^{t}_{c}{}_{t}$ from the classical collapse) at $t \approx 0.94$.}\label{fig:difference}
\end{figure}

To compare to classical gravitational collapse, we have also performed the corresponding simulations in the purely classical theory, with the initial data of $\Phi$ in the classical simulations matching the initial data of $\langle\hat\Phi\rangle$ in the quantum ones. On the scale of the plots in Fig.~\ref{fig:fig} one is not able to distinguish the classical from quantum results. Given that, in Fig.~\ref{fig:difference}, we extract the quantum corrections of $\langle \hat{T}^{t}{}_{t}\rangle$ ({\it i.e.}, the difference between $\langle \hat{T}^{t}{}_{t} \rangle$ of the quantum simulation and the corresponding classical $T^{t}_{c}{}_{t}$) at $t \approx 0.94$ in the case where a black hole forms. To assess the reliability of these corrections, we perform a convergence study for various numbers of quantum modes used. We see that the quantum corrections already converge at $N_{\rm mode}=4900$.

{\bf Summary~} 
\label{sec:appendix}
We have developed a formalism to simulate fully quantum fields collapsing to form a black hole. In this formalism, by choosing a convenient coherent quantum state, the equation of motion for the expectation value of the field operator coincides with that of the classical field in classical numerical relativity simulations, which means that all previous numerical relativity methods may be applied to the evolution of fully quantum fields by additionally attaching the evolution of the quantum modes. For free fields, including fermionic fields and U(1) gauge fields, all the mode functions can be evolved, as explicitly shown in this paper, while for interacting quantum fields some approximations are needed for the evolution of the quantum fields, such as the Hartree approximation \cite{Borsanyi:2007wm,Saffin:2014yka}. We have focused on spherical symmetry in our simulations, which is computationally much less expensive, but $1/r^n$-type terms appear in the equations of motion, which require extra care to handle in numerical schemes. The formalism can be applied to full 4D simulations, whose equations of motion are even free of the difficult $1/r^n$ terms, but they require considerably more computing resources. Detailed investigations of the quantum effects of gravitational collapse such as the presence of Hawking radiation and corrections to critical collapse will be discussed in a future work \cite{futurepaper}. 

\section*{Acknowledgements} 
\label{sec:acknowledgements}
BB is supported by an STFC studentship, PMS acknowledges support from STFC grant ST/P000703/1 and SYZ acknowledges support from the starting grants from University of Science and Technology of China under grant No.~KY2030000089 and GG2030040375 and is also supported by National Natural Science Foundation of China under grant No.~12075233 and 11947301.

\bibliography{sample.bib}

\end{document}